\documentclass[twocolumn]{aastex631}

\usepackage{natbib}
\usepackage{hyperref}
\usepackage{appendix}

\begin{document}

\title{The High-Redshift Gas-Phase Mass-Metallicity Relation in FIRE-2}

\author[0000-0001-6676-4132]{Andrew Marszewski\footnotemark}
\affiliation{CIERA and Department of Physics and Astronomy, \\
Northwestern University, \\
1800 Sherman Ave, 8th Floor CIERA, Evanston, IL 60201}

\author[0000-0003-4070-497X]{Guochao Sun}
\affiliation{CIERA and Department of Physics and Astronomy, \\
Northwestern University, \\
1800 Sherman Ave, 8th Floor CIERA, Evanston, IL 60201}

\author[0000-0002-4900-6628]{Claude-André Faucher-Giguère}
\affiliation{CIERA and Department of Physics and Astronomy, \\
Northwestern University, \\
1800 Sherman Ave, 8th Floor CIERA, Evanston, IL 60201}

\author[0000-0003-4073-3236]{Christopher C. Hayward}
\affiliation{Center for Computational Astrophysics, \\ 
Flatiron Institute, \\ 
162 Fifth Avenue, New York, NY 10010}

\author[0000-0002-1109-1919]{Robert Feldmann}
\affiliation{Department of Astrophysics, \\ 
University of Zurich, \\
Zurich CH-8057, Switzerland}

\begin{abstract}
\footnotetext{Corresponding Author: Andrew Marszewski \\ \href{mailto:AndrewMarszewski2029@u.northwestern.edu}{AndrewMarszewski2029@u.northwestern.edu}}
The unprecedented infrared spectroscopic capabilities of JWST have provided high-quality interstellar medium (ISM) metallicity measurements and enabled characterization of the gas-phase mass-metallicity relation (MZR) for galaxies at $z \gtrsim 5$ for the first time.  We analyze the gas-phase MZR and its evolution in a high-redshift suite of FIRE-2 cosmological zoom-in simulations at $z=5-12$ and for stellar masses $M_* \sim 10^6-10^{10} \rm{M}_\odot$. These simulations implement a multi-channel stellar feedback model and produce broadly realistic galaxy properties, including when evolved to $z=0$. The simulations predict very weak redshift evolution of the MZR over the redshift range studied, with the normalization of the MZR increasing by less than $0.01$ dex as redshift decreases from $z = 12$ to $z=5$.  The median MZR in the simulations is well-approximated as a constant power-law relation across this redshift range given by $\log(Z/Z_\odot) = 0.37\log(M_*/\rm{M}_\odot) - 4.3$. We find good agreement between our best-fit model and recent observations made by JWST at high redshift. The weak evolution of the MZR at $z > 5$ contrasts with the evolution at $z \lesssim 3$, where increasing normalization of the MZR with decreasing redshift is observed and predicted by most models.  The FIRE-2 simulations predict increasing scatter in the gas-phase MZR with decreasing stellar mass, in qualitative agreement with some observations.
\end{abstract}


\section{Introduction} \label{sec:intro}

The mass-metallicity relation (MZR) is the observed positive correlation between a galaxy’s stellar mass and its metallicity (\citealp{1979A&A....80..155L, Tremonti_2004}).  There are both stellar and gas-phase versions of the MZR, which relate stellar mass to stellar metallicity and interstellar medium (ISM) metallicity, respectively.  Throughout this work we focus on the gas-phase MZR.  The MZR and its evolution have been observed extensively across wide ranges of redshift and stellar mass (e.g., \citealp{Erb_2006, Lee_2006, Zahid_2011, Zahid_2012, Henry_2013a, Henry_2013b, Maier_2014, Steidel_2014, Yabe_2014, Sanders_2015, Guo_2016}). \citet{Zahid_2013} characterized the observed evolution of the MZR from $z=0-2.3$, noting that, for a given stellar mass, metallicity tends to increase as redshift decreases. Previously, relatively small samples of galaxies have been able to confirm the existence of the MZR up to $z \sim 3$ (e.g., \citealp{Maiolino_2008, Mannucci_2009}).

Recently, new observations from the James Webb Space Telescope (JWST) have greatly expanded the physical regimes where the MZR has been probed, both in mass and in cosmic time.  For example, \citet{Nakajima_2023} characterize the evolution of the MZR for $4<z<10$ using metallicity measurements of 135 galaxies identified by JWST in this redshift range.  \citet{Curti_2023b} analyze the gas-phase metallicities of 146 high-redshift ($3<z<10$) galaxies observed by JWST, 80 of which were also present in the sample from \cite{Nakajima_2023}.  \citet{Bunker_2023} use strong-line ratios to constrain the metallicity of GN-z11 at $z \sim 10.6$.  Gas-phase metallicities have been derived for a number of other high-redshift JWST targets via direct methods (e.g., MACS0647-JD at $z = 10.165$; \citealp{Hsiao_2024}, galaxies in JWST Early Release Observations at $z \sim 8$; \citealp{Curti_2023a}, and 9 sources in the sight line of MACS J1149.5+2223 at $z = 3-9$; \citealp{Morishita_2024}).  It is imperative that these unprecedented advances in observations of the MZR at high redshift and low stellar mass be met with detailed theoretical predictions in the newly probed regimes.

The MZR and its evolution have been studied at high redshift in a number of different simulation codes, such as IllustrisTNG \citep{Torrey_2019}, FirstLight \citep{Langan_2020}, SERRA \citep{Pallottini_2022}, ASTRAEUS \citep{Ucci_2023}, and FLARES \citep{Wilkins_2023}. 
 FirstLight, ASTRAEUS, and FLARES predict weak or no evolution in the MZR for $z \gtrsim 5$.  The Feedback in Realistic Environment (FIRE) project\footnote{See FIRE project website: \url{http://fire.northwestern.edu}.} is a set of cosmological zoom-in simulations that resolve the multiphase ISM of galaxies and implement detailed models for star formation and stellar feedback (\citealp{Hopkins_2014,Hopkins_2018, Hopkins_2023}).  \citet{Ma_2016} characterized the MZR in the first generation of FIRE simulations from $z=0$ to $z=6$.  A number of previous works (e.g., \citealp{Ma_2016, Torrey_2019, Langan_2020}) invoke gas fractions to explain the redshift evolution or lack of redshift evolution in the MZR.  Recently, \citet{Bassini_2023} analyze the evolution of the MZR for $z=0-3$ in FIREbox \citep{Feldmann_2023}, a cosmological volume simulation that uses FIRE-2 physics.  

In this work, we present the MZR predicted from a high-redshift suite of FIRE-2 simulations.  We measure the MZR at redshifts $5\leq z\leq12$ and we provide fitting formulae describing the redshift evolution of the MZR.  We show that our model is in agreement with recent high-redshift observations of the MZR made by JWST.  We compare our best-fit MZR with previous simulation-based models across this redshift range. This work analyzes the same suite of FIRE-2 simulations as done in two recent papers by \citet{Sun_2023a,Sun_2023b}, which focus on bursty star formation and its implications for the high-redshift ultraviolet luminosity function (UVLF) and survey selection effects in the context of JWST.  \citet{Yang_2023} also analyzed some galaxies from this suite of simulations and showed that metallicities derived from mock observations of emission lines from individual HII regions of FIRE-2 galaxies are in $1\sigma$ agreement with JWST and ALMA observations.

This paper is organized as follows.  In Section \ref{sec:methods}, we describe the high-$z$ suite of FIRE-2 simulations used in this paper and the methods used to analyze the gas-phase MZR.  In Section \ref{sec:results}, we present the resulting high-$z$ MZR in FIRE-2 simulations.  We compare our results with new observations made by JWST and with previous theoretical models of the MZR derived from FIRE-1 and other simulations.  Finally, in Section \ref{sec:Conclusions} we summarize the key conclusions from this work and discuss potential future work on high-$z$ metallicity scaling relations.

Throughout this work we adopt a standard flat $\Lambda$CDM cosmology with cosmological parameters consistent with \citet{Planck_2020}.  We define a galaxy's gas-phase metallicity to be the mass-weighted mean metallicity of all gas particles within $0.2R_{\rm vir}$ of the galaxy's center.  All $\log$ functions are base 10, except when written as $\ln$ (natural logarithm).  

\section{Methods} \label{sec:methods}

\subsection{Simulations}

The simulations analyzed in this paper are cosmological zoom-in simulations from a FIRE-2 high-redshift suite originally presented by \citet{Ma_2018a,Ma_2018b,Ma_2019}.  All simulations in this suite were run using the GIZMO code \citep{Hopkins_2015}.  The hydrodynamic equations are solved using GIZMO's meshless finite-mass (MFM) method.  The 34 particular simulations analyzed in this paper are the z5m12a--e, z5m11a--i, z5m10a--f, z5m09a--b, z7m12a--c, z7m11a--c, z9m12a, and z9m11a--e runs.  The names of these simulations denote the final redshift that they were run down to ($z_{\rm fin}=5$, $7$, or $9$) and the main halo masses (ranging from $M_{\rm h} \approx 10^9-10^{12} \rm{M}_\odot$) at these final redshifts.  Baryonic (gas and star) particles have initial masses $m_b = 100-7000 \rm{M}_\odot$ (simulations with more massive host galaxies have more massive baryonic particles).  Dark matter particles are more massive by a factor of $\Omega_{DM}/\Omega_b \approx 5$. Gravitational softenings are adaptive for the gas (with minimum Plummer-equivalent force softening lengths $\epsilon_b = 0.14–0.42$ physical pc) and are fixed to $\epsilon_* = 0.7–2.1$ physical pc and $\epsilon_{DM} = 10-42$ physical pc for star and dark matter particles, respectively.

A full description of the baryonic physics in FIRE-2 simulations is given by \citet{Hopkins_2018}, while more details on this specific suite of FIRE-2 simulations are discussed in \citet{Ma_2018a,Ma_2018b,Ma_2019}.  Here, we briefly review the aspects of the simulations most pertinent to our MZR analysis.  

FIRE-2 simulations track the abundances of 11 different elements (H, He, C, N, O, Ne, Mg, Si, S, Ca, Fe).  Metals are returned via multiple stellar feedback processes, including core collapse and type Ia supernovae as well as winds from O/B and AGB stars.  Star particles in the simulations represent stellar populations with a Kroupa IMF \citep{Kroupa_2001} and with the stellar evolution models from STARBURST99 \citep{Leitherer_1999}.  These simulations also include sub-grid modelling for turbulent diffusion of metals to allow for chemical exchange between neighboring particles.  The implementation and effects of the sub-grid turbulent diffusion model in FIRE simulations are described in \citet{Colbrook_2017} and \citet{Escala_2018}.

\subsection{Analysis}

\begin{figure*}[t!]
    \centering
    \includegraphics[width=\linewidth]{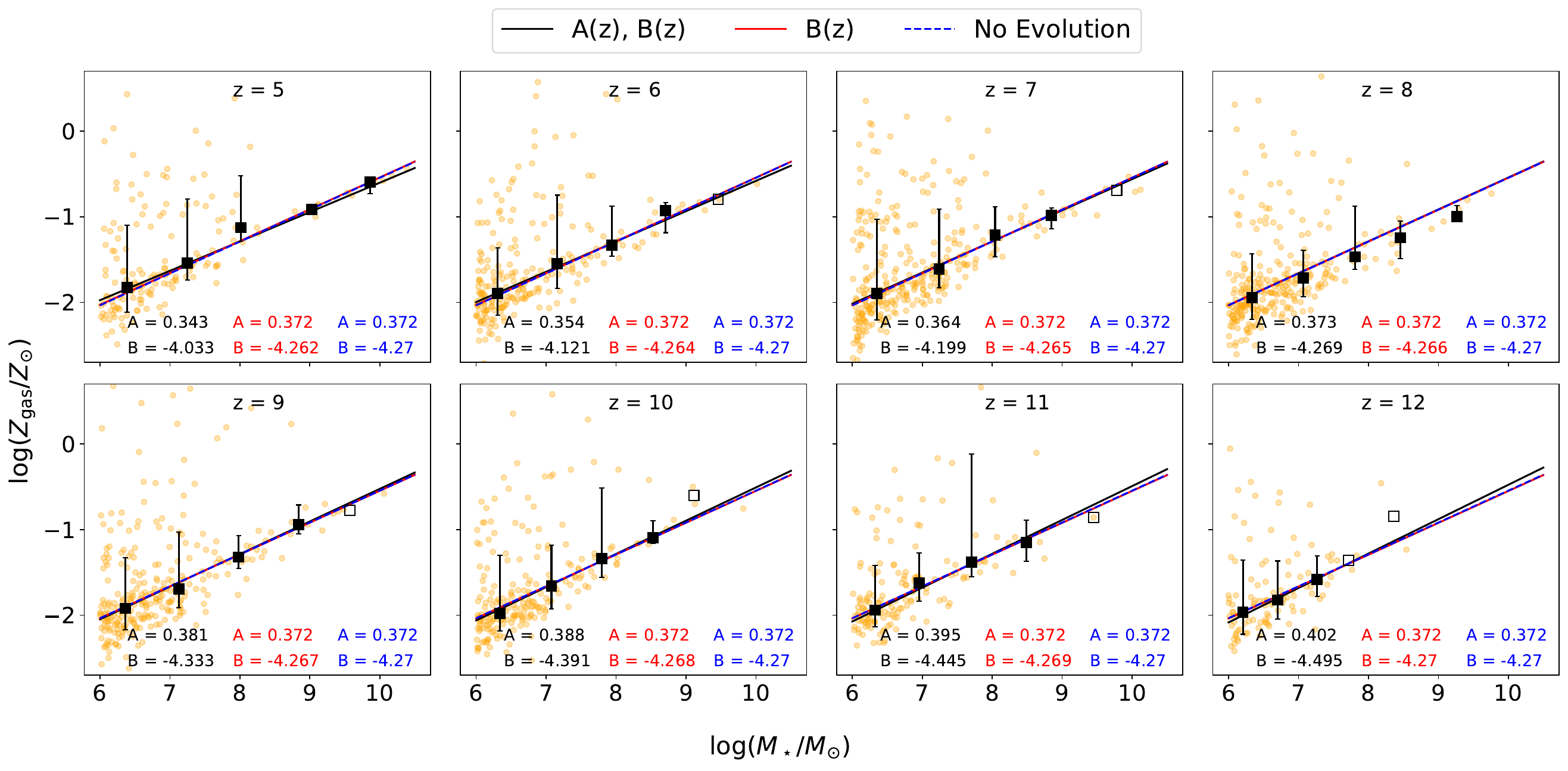}
    \caption{The evolution of the gas-phase MZR in FIRE-2 simulations from $z = 5-12$.  The solid black, solid red, and dashed blue lines show our best fits to the medians for the Slope and Normalization Evolution, Normalization Evolution, and No Evolution models, respectively.  Values for the slope ($A$) and normalization ($B$) are provided for each model at the bottom right of each redshift panel in their corresponding color.  Metallicities for individual galaxies are shown in orange.  The median metallicities of stellar mass bins for stellar mass bins containing at least five galaxies are shown by solid black squares with error bars representing the 16th and 84th percentiles.  Empty black squares represent the median metallicities of stellar mass bins that contain fewer than 5 galaxies.  Both the slope and normalization of the MZR are approximately constant over this redshift interval.}
    \label{fig:MZR}
\end{figure*}

We analyze galaxies in each simulation at snapshots from $z=5$ to $z=12$, with integer redshift increments.  In addition to the main, most massive galaxy, each simulation's zoom-in region captures numerous other, less massive, galaxies.  The coordinates of galaxy centers and virial radii are taken from Amiga Halo Finder (AHF) catalogs \citep{Gill_2004,Knollmann_2009}.  Halos are defined using the redshift-dependent overdensity parameter from \citet{Bryan_1998}.  The galaxies used in our analysis are filtered based on the following criteria.  Galaxies must have a minimum stellar mass within $0.2R_{\rm vir}$ $M_{*}>10^{6}\rm{M}_\odot$, a non-zero gas mass ($M_{\rm gas}>0$) within the same radius, and the halo must have a minimum virial mass given by $M_{\rm vir} \geq 10^{9}\rm{M}_\odot$.  We exclude satellite galaxies and subhalos from our analysis as their properties can be significantly influenced by their host galaxy.  Finally, we filter out any galaxies that are “contaminated" by low-resolution dark matter particles residing within $1R_{\rm vir}$ of their center.  After applying these cuts, the number of galaxies in our sample at different redshifts ranges from $N_{\rm gal} = 106-300$.

\begin{table*}[t!]
\centering
\begin{tabular}{c c c c c} 
 \hline
 Model & $a$ & $b$ & $\alpha$ & $\beta$ \\
 \hline
 Slope and Normalization Evolution (A(z), B(z)) & 0.3717 & $-4.251$ & 0.2065 & 0.1493 \\ 
 Normalization Evolution (B(z)) & 0.3683 & $-4.232$ & 0 & 0.0380 \\
 No Evolution & 0.3702 & $-4.241$ & 0 & 0 \\
 \hline
\end{tabular}
\caption{Best fit parameters for three different evolution models of the MZR using the form given by equation \ref{eqn:Fit}.  For the first model, both the slope and normalization of the MZR are allowed to vary smoothly with $(1+z)$.  The second model fixes the slope of the fit but allows for the normalization to vary.  The third model fixes both the normalization and slope across our entire redshift range ($z=5-12$).}
\label{table:1}
\end{table*}

This work focuses on the MZR for gas-phase metallicity, which reflects the current chemical composition of a galaxy's ISM.  We leave the analysis of stellar metallicities, which provide information on the integrated chemical enrichment history of galaxies, as a subject for future work.  We define a galaxy's gas-phase metallicity to be the mass-weighted mean metallicity of all gas particles within $0.2R_{\rm vir}$ of the galaxy's center.  We calculate a galaxy's stellar mass as the total mass of all star particles within $0.2R_{\rm vir}$ of the its center.  We choose $0.2R_{\rm vir}$ as the outer boundary for our galaxies rather than the commonly used $0.1R_{\rm vir}$ due to the tendency of high-redshift galaxies to have more expansive stellar populations relative to their virial radii as compared to galaxies at lower redshift.  This is consistent with the radial cut used by \citet{Sun_2023a,Sun_2023b} in their analysis of the same simulation suite.  

In Appendix \ref{appendix:Gas_Cuts} we consider alternative definitions and gas cuts for calculating gas-phase metallicities.  This includes weighting by gas particles' SFR property (rather than by mass) when calculating metallicity, setting the outer boundary of galaxies to be $0.1R_{\rm vir}$ (rather than $0.2R_{\rm vir}$), and introducing a temperature cut of $T_{\rm gas} < 10^4$ K on gas particles used to calculate metallicity.  The SFR-weighting scheme has only a minor impact on our calculated MZR and likely introduces a bias by removing galaxies with no star-forming gas from our sample since, according to the fundamental metallicity relation, galaxies with lower star formation rate will tend to have higher metallicities at fixed stellar mass (\citealp{Ellison_2008, Mannucci_2010, Marszewski_inprep}).  We also show that our MZR is relatively insensitive to the different radial and temperature cuts applied to gas particles in our analysis. 

\section{Results} \label{sec:results}

In this section we present the MZR for our analyzed galaxies.  Gas-phase metallicities are given in units of $\log(Z/Z_\odot)$, where $Z$ is the total metal mass fraction and we have adopted the solar metallicity, $Z_\odot = 0.02$, from \citet{Anders_1989}.  With this value of $Z_\odot$, we can convert to units of oxygen abundance (often more relevant for comparing with observations) using the calibration presented in Appendix B of \citet{Ma_2016}, 
\begin{equation}
    \log(Z/Z_\odot) = 12 + \log(\rm{O}/\rm{H}) - 9.00.
\end{equation}
This calibration was obtained by fitting metallicity against oxygen abundance in FIRE-1 simulated galaxies and may be subject to systematic uncertainties originating from supernovae rates, metal yields from different enrichment processes, and the fiducial solar metallicity used in FIRE simulations.

\begin{figure*}[t!]
    \centering
    \includegraphics[width=\linewidth]{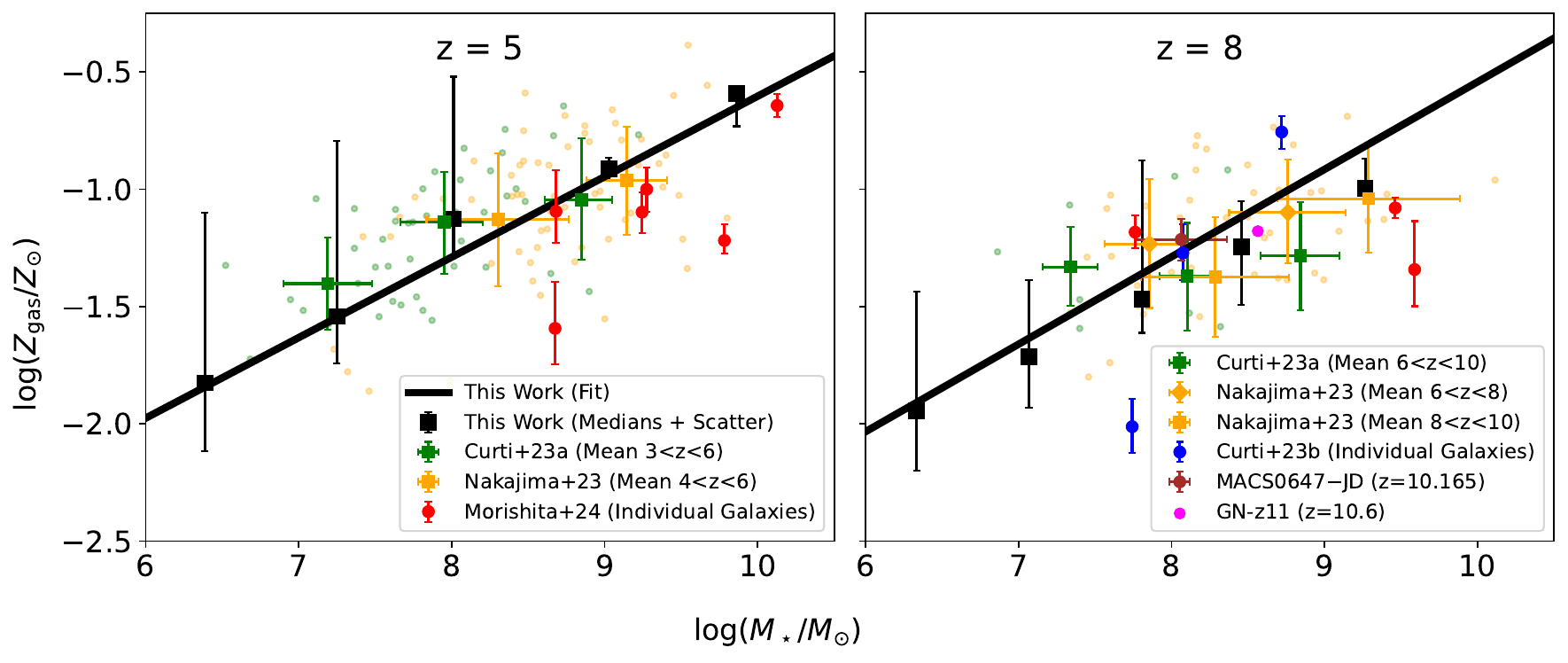}
    \caption{Comparison between our best-fit and recent high-redshift observations of the MZR made by JWST.  We center our comparisons at $z=5$ (left panel) and $z=8$ (right-panel).  As in Figure \ref{fig:MZR}, solid black squares show the median metallicities of stellar mass bins with error bars representing the 16th and 84th percentiles.  Our model shows good agreement with the stellar-mass-binned mean values from both \citet{Nakajima_2023} (orange squares and diamonds) and \citet{Curti_2023b} (green squares).  Measurements of individual galaxies are represented by dots, including 3 JWST Early Release Observations at $z \sim 8$ presented by \citet{Curti_2023a} (shown in blue), 9 measurements at $z = 3-9$ presented by \citet{Morishita_2024} (shown in red), the MACS0647-JD galaxy presented by \citet{Hsiao_2024} at $z \sim 10.16$ (shown in brown), and the GN-z11 galaxy presented by \citet{Bunker_2023} at $z \sim 10.6$ (shown in magenta).}
    \label{fig:Observations}
\end{figure*}

\subsection{Mass-Metallicity Relation}

At each redshift analyzed, we separate the data into 5 stellar mass bins of equal width. These stellar mass bins span from the minimum stellar mass of galaxies included in our sample ($M_{*}=10^{6}\rm{M}_\odot$) to the maximum stellar mass of galaxies in our sample at that redshift.  We then calculate linear best fit models between the median values of $\log(Z/Z_\odot)$ and $\log(M_*/M_\odot)$ in different stellar mass bins.  For purposes of fitting, the median metallicity value of each stellar mass bin is given equal weight.  We provide a fitting formula across our redshift range ($z \in [5,12]$) by allowing the slope and y-intercept of the linear fit to vary with $(1+z)$, using the form:
\begin{equation} \label{eqn:Fit}
    \log(Z/Z_\odot) = A(z) \log(M_*/M_\odot) + B(z), 
\end{equation}
where
\begin{equation}
    A(z) = a\left(\frac{1+z}{1+8}\right)^\alpha,
\end{equation}
\begin{equation}
    B(z) = b\left(\frac{1+z}{1+8}\right)^\beta.
\end{equation}

To investigate the importance of the redshift evolution of the slope and normalization of our fits, we also provide fitting formulae where we fix combinations of these parameters across our redshift range. In particular we present one version of the fit where we fix the slope of the MZR by setting $\alpha = 0$ and another version of the fit where we fix both the slope and normalization by setting $\alpha = \beta = 0$.  Our best-fit parameters for all three versions of the fit are given in Table \ref{table:1}.  The MZR for each analyzed galaxy along with the medians of each stellar mass bin and the three versions of the best-fit lines are plotted for all analyzed redshifts in Figure \ref{fig:MZR}.

The agreement between our redshift-evolving fits and our fit with no evolution suggests weak redshift evolution of the MZR across our redshift range.  Our Slope and Normalization Evolution fit is characterized by a decrease in slope and increase in normalization of the MZR with decreasing redshift.  The effects of these changes, however, largely offset one another across our stellar mass range.  From the Normalization Evolution version of our fit we find that, when holding the slope constant, the normalization of the MZR changes by only $\approx 0.01$ dex over our entire analyzed redshift range ($z = 5-12$).  The evolution found in either of these models is within the level of inherent uncertainty in our data, as evidenced by their agreement with the No Evolution model.  We therefore conclude that the MZR in our simulations is characterized by weak to no redshift evolution for $z \gtrsim 5$.  

We also find that the scatter of our relation tends to decrease with increasing stellar mass.  For example, our lowest two stellar mass bins, centered below $10^{7.5} M_\odot$, have scatters between $0.6-1.2$ dex, while stellar mass bins centered above $10^{8.5} M_\odot$ typically have scatter less than $0.5$ dex.  Therefore, FIRE-2 simulations predict large scatter in the MZR at low stellar mass.  We hypothesize that this is, in part, due to galaxies becoming less bursty relative to their stellar mass as their stellar mass increases.  This explanation would be consistent with the fundamental metallicity relation (FMR) where the star-formation rate (or gas-fraction) acts as a secondary predictor for metallicity (\citealp{Ellison_2008,Mannucci_2010}).  According to the FMR, smaller variance in star-formation rates at a given stellar mass would result in smaller scatter in the metallicities at that stellar mass.  The scatter we predict at the low-mass end is much larger than predicted by \citet{Bassini_2023} in their analysis of FIREbox simulations at $z \leq 3$, but it is not inconsistent with that study as the mass range where we predict increased scatter is below the mass limit of their analysis ($M_{*} \sim 10^{8}$ M$_{\odot}$). 
There is also likely some redshift evolution of the scatter between $z=5$ and $z=3$. We verified using FIRE-2 zoom-in simulations evolved to $z = 0$ that the scatter in the MZR at $z = 3$ increases significantly below the lower stellar mass limit used by \citet{Bassini_2023} but is in agreement with their results above this limit.  We do not find a clear redshift evolution trend of the scatter of the MZR at fixed stellar mass over our redshift range in FIRE-2.

\subsection{Comparison with Observations}

The advent of JWST has allowed for the measurement of significant samples of ISM metallicities at $z \geq 5$ for the first time.  JWST surveys have already made many metallicity measurements available for galaxies at much earlier redshift than previously feasible (e.g., \citealp{Curti_2023b, Nakajima_2023}). 
 \citet{Hsiao_2024} measure the metallicity of MACS0647-JD using the direct, $T_e$-based method at $z = 10.165$ and \citet{Bunker_2023} present a metallicity measurement of the galaxy GN-z11 using strong-line ratios at $z = 10.6$, further demonstrating the observational power of JWST.  With additional large samples of metallicity measurements from JWST on the way, it is timely that we verify the results of our simulations against current observations and make predictions for future observations.  Here, we compare our best-fit MZR to observations already made available from JWST surveys.  

 \begin{figure*}[t!]
    \centering
    \includegraphics[width=\linewidth]{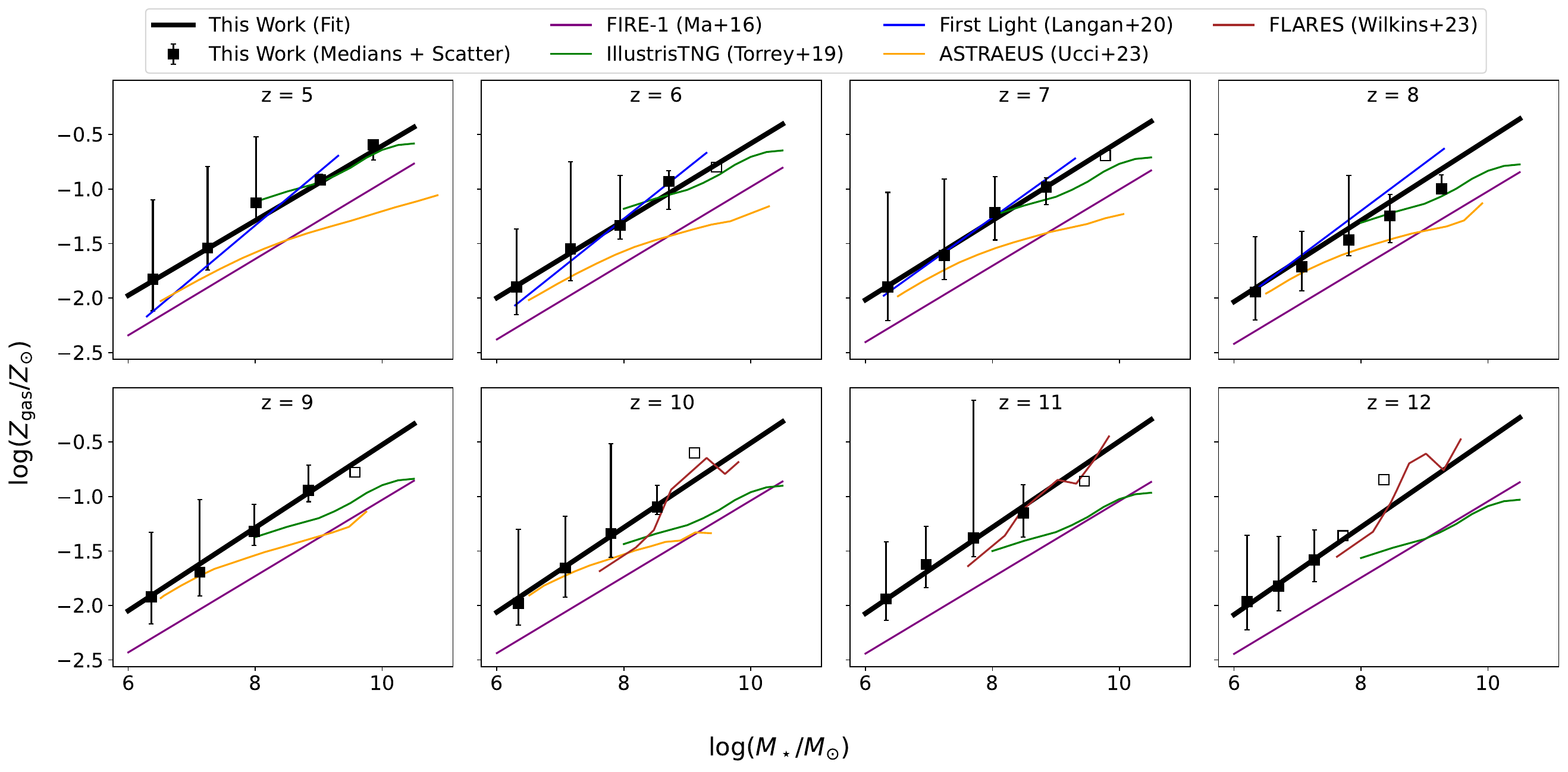}
    \caption{Comparison between our best-fit model and previous simulation-based work on the MZR.  Results are shown for IllustrisTNG \citep{Torrey_2019}, FirstLight \citep{Langan_2020}, ASTRAEUS \citep{Ucci_2023}, FLARES \citep{Wilkins_2023} for the redshift and stellar mass ranges at which they were reported.  Additionally, the MZR fit for FIRE-1 data from \citet{Ma_2016}, given in the range $z = 0-6$, has been extrapolated and plotted across our redshifts of interest.  With the exception of IllustrisTNG, all models shown here are consistent with weak evolution of the MZR over this redshift interval.}
    \label{fig:Models}
\end{figure*}

\citet{Nakajima_2023} and \citet{Curti_2023b} present metallicity measurements made by JWST's Near Infrared Spectrograph (NIRSpec) Instrument of 135 and 146 galaxies, respectively, primarily using strong line methods.  There are 80 galaxies that are originally presented by \citet{Nakajima_2023} which are also included in the analysis by \citet{Curti_2023b}.  The sample from \citet{Nakajima_2023} includes 10 direct, $T_e$-based measurements.  \citet{Curti_2023b} include 3 direct, $T_e$-based measurements, originally presented by \citet{Curti_2023a}, as well as the measurement of GN-z11 by \citet{Bunker_2023}, all of which we compare to individually.  Galaxies analyzed in \citet{Curti_2023b} have redshifts in the range $3 < z < 10$ and stellar masses in the range $10^{6.5} < M_{*}/M_\odot < 10^{10}$ while galaxies in \citet{Nakajima_2023} have redshifts $4 < z < 10$ and stellar masses $10^7 < M_{*}/M_\odot < 10^{10}$.  Each stellar mass-binned mean metallicity from these two works is in reasonably good agreement with our best-fit MZR.  However, both works report a smaller slope in the observed MZR ($A = 0.17 \pm 0.03$ from \citealp{Curti_2023b} and $A = 0.25 \pm 0.03$ from \citealp{Nakajima_2023}).  \citet{Morishita_2024} provide 9 additional gas-phase metallicity measurements of galaxies with redshifts $3 < z < 9$ made via the direct method.  A comparison between the best-fit model presented in this work and recent high-z JWST observations of the MZR is shown in Figure \ref{fig:Observations}.  Additionally, \citet{Curti_2023b} report a significant difference in normalization between the high-redshift MZR and the MZR in the local universe.  The MZR is found to be offset by an average of $-0.48$ dex and $-0.64$ dex in the $z=3-6$ and $z=6-10$ redshift bins, respectively, as compared to the low-mass extrapolation of the MZR in the local universe presented by \citet{Curti_2020}. 
 These findings support the notion that the MZR evolves significantly for $z \lesssim 3$ and evolves weakly or not at all for $z \gtrsim 3$.  \citet{Nakajima_2023} do not find significant evidence for evolution of the MZR over the redshift range $z=4-10$.

 Increasing scatter of the MZR with decreasing stellar mass is found in several observational works at lower redshift (e.g., \citealt{Zahid_2012} at $z \leq 0.1$, \citealt{Guo_2016} at $0.5 \leq z \leq 0.7$), qualitatively matching the trend we find in FIRE-2 at higher redshift.  More recently, \citet{Li_2023} find an increase in the scatter at the low-mass end of the MZR using a sample of 51 dwarf galaxies observed by JWST at $z=2-3$.  They quote intrinsic scatters in their MZR of $0.16-0.18$ dex and $0.23$ dex for stellar mass bins at $10^8-10^9 M_\odot$ and $10^7 M_\odot$, respectively.

 As shown in Figure \ref{fig:Observations}, there are apparent quantitative inconsistencies between the scatter predicted by our model and the scatter observed by \citet{Curti_2023b} and \citet{Nakajima_2023}.  At lower stellar masses ($M_{*}/M_\odot \lesssim 10^{8.5}$), the scatter in the simulations is larger than that of the observations.  In part, this may be a result of observational samples at high redshift having a selection bias toward more luminous, actively star-forming galaxies \citep[e.g.,][]{Sun_2023a}.  According to the fundamental metallicity relation, galaxies with low star formation rates that are absent from observational samples would have systematically higher metallicites at a given stellar mass.  This explanation is consistent with our analysis since the increased scatter in our MZR is largely due to a small number of very metal rich galaxies.  Additionally, this could be a result of strong line measurements of metallicity, which constitute the majority of these observational samples, systematically underestimating the scatter in the MZR.  Strong line measurements are susceptible to predicting systematically low scatter since they rely on ratios between specific strong emission lines and do not take into account potential scatter in other parameters used to infer metallicities from these ratios.  For example, it is possible for two different galaxies to have different metallicities yet have the same line ratio (e.g, if the ionization state of the gas is different).  A particular strong line calibration would then imply a single metallicity.  At higher stellar masses ($M_{*}/M_\odot \gtrsim 10^{8.5}$), where the predicted MZR is very narrow, the scatter is smaller in simulations than in observations.  This could be because of the uncertainties present in observational methods (e.g., noise in raw observational data and various uncertainties in converting from line fluxes to metallicities) that are not present in the analysis of simulations.  These uncertainties could result in a larger apparent scatter of the MZR in a regime where the relation is particularly tight in nature.

\subsection{Comparison with Other Theoretical Models}

Some other cosmological and seminumerical simulation projects have analyzed the MZR at high redshift.  We compare our work on the MZR in high-redshift FIRE-2 simulations to other theory-based work and discuss predictions for future observations.  Figure \ref{fig:Models} shows comparisons between the best-fit model presented in this work and other theoretical and simulation-based models.

The weak time evolution of the MZR we find at $5 \leq z \leq 12$ is consistent with previous results from FIRE-1 simulations reported by \citet{Ma_2016}, who predicted a flattening of the evolution of the MZR at high redshift.  However, our best-fit model predicts metallicities approximately $0.3$--$0.4$ dex higher than their best fit derived from FIRE-1 simulations.  \citet{Bassini_2023} find significant evolution in the MZR from $z=0-3$ in the FIREbox simulation, a full cosmological volume simulation that uses the physics of FIRE-2.  In particular, the gas-phase metallicity at fixed stellar mass is found to increase with decreasing redshift over the range $z=0-3$.  Our best-fit MZR at $z=5$ is similar in normalization and slope to the FIREbox results at $z=3$.  We therefore conclude that the evolution of the MZR in FIRE-2 simulations is characterized by metallicity increasing with decreasing redshift for $z \lesssim 3$ and by little to no evolution for $z \gtrsim 3$ at fixed stellar mass.  A more complete comparison between this work and previous work on the MZR in FIRE simulations is provided in Appendix \ref{appendix:FIRE}.

Weak evolution of the MZR past $z \gtrsim 5$ has been found in some other simulation projects as well.  \citet{Langan_2020} report a slight increase in mean metallicity with increasing redshift in FirstLight simulations from $z = 5-8$.  They do not, however, find this trend to be significant at a level beyond the intrinsic scatter of their data.  Similarly, \citet{Ucci_2023} characterize the MZR in ASTRAEUS, a seminumerical simulation project, as having effectively no redshift evolution from $z=5-10$.  The median metallicity of their low-mass galaxies ($M_* \sim 10^7 M_\odot$) decreases very slightly ($\sim 0.15$ dex) from $z=10$ to $z=5$, while metallicities in their high-mass range ($M_* \sim 10^9 M_\odot$) remain nearly constant over the redshift range.  \citet{Wilkins_2023} also find no significant evolution in the MZR for $z \geq 10$ in FLARES.  \citet{Torrey_2019} report that, for IllustrisTNG simulations from $z = 2-10$, the normalization of the MZR decreases with increasing redshift while the slope is not a strong function of redshift.  The evolution of the normalization in their simulations within their redshift range is given by $d\log(Z)/dz \approx -0.064$.  This evolution is stronger than the evolution found in FIRE-2 and  other simulation projects.

\subsection{Interpretation using Analytic Models}

Multiple explanations for the weak evolution of the MZR beyond $z \gtrsim 5$ have been put forth.  \citet{Torrey_2019} find that the evolution of the MZR from $z=2$ to $z=10$ is explained by the evolution of the gas fraction through the gas-regulator model.  The regulator model gives an approximate equilibrium metallicity of the form (e.g., \citealp{Lilly_2013})
\begin{equation} 
    Z_{\rm eq} = Z_{\rm acc} + \frac{y}{1+(1-R)^{-1}\eta+f_{\rm gas}},
\end{equation}
where $y = M_{\rm Z}/M_\star$ (often assumed to be $y=0.02$) is the metal yield (the mass of metals returned to the ISM per unit mass in formed, long-lived stars), $Z_{\rm acc}$ is the metallicity of accreted gas, $\eta = \dot{M}_{\rm wind} / \rm{SFR}$ is the mass loading factor of galactic winds, where $\rm{SFR}$ is a galaxy's star formation rate, and $f_{\rm gas}$ is the version of the galactic gas fraction given by $M_{\rm gas}/M_\star$.  However, \citet{Bassini_2023} show that the evolution of the gas fraction does not drive the evolution of the MZR in FIREbox at lower redshifts ($z=[0,3]$).  Rather, an evolution in both the mass-loading factor and the metallicities of inflows and outflows at fixed stellar mass drive the decrease of the MZR with increasing redshift up to $z=3$.  In Appendix \ref{appendix:Gas_Fraction} we show there is substantial redshift evolution in the median gas fractions of our analyzed galaxies with redshift.  This fact combined with the very weak evolution of the MZR over the same redshift range implies that the weak evolution of the high-redshift MZR is likely not explained solely by the evolution of the gas fraction in the gas-regulator model.  

Other works have used idealized “closed box" or “leaky box" models to explain the weak evolution of the MZR at high redshift (e.g., \citealp{Langan_2020, Ma_2016}).  In the “closed box" model, metallicity is given by  
\begin{equation} 
    Z_{\rm gas} = -y \ln({\tilde{f}_{\rm gas}}),
\end{equation}
while in the “leaky box" model,
\begin{equation} 
    Z_{\rm gas} = -y_{\rm eff} \ln({\tilde{f}_{\rm gas}}),
\end{equation}
where $\tilde{f}_{\rm gas}$ is the version of the gas fraction given by $M_{\rm gas}/(M_\star + M_{\rm gas})$ and $y_{\rm eff}$ is the effective metal yield which is often calibrated to make the “leaky box" model best fit the data.  In Appendix \ref{appendix:Gas_Fraction} we present median values of $\tilde{f}_{\rm gas}$ across our redshift range in different stellar mass bins.  The difference between $y$ and $y_{\rm eff}$ quantifies the net impact of inflows and outflows on a galaxy's metallicity.  From this picture it has been argued that the weak evolution of the high-redshift MZR is due to values of $\tilde{f}_{\rm gas}$ that have saturated to unity and/or that evolve weakly at high redshift.  \citet{Langan_2020} find that a “leaky box" model with $y_{\rm eff} = 0.002$ is able to explain the weak evolution of the MZR predicted by FirstLight simulations.  However, with an effective yield that is an order of magnitude lower than the intrinsic stellar yield ($y \approx 0.02$), this model concedes that the impact of inflows and outflows is crucial.  \citet{Ma_2016} also found that a “closed box" model systematically over-predicts metallicities in FIRE-1 simulations, also due to the large effects of inflows and outflows (i.e. $y_{\rm eff}$ is significantly smaller than $y$).  Thus, weakly evolving gas fractions are at best an incomplete explanation for the weak evolution of the MZR at high redshift. It would be interesting to perform an analysis similar to that done by \citet{Bassini_2023} to quantify explicitly the effects of gas fractions, inflow/outflow metallicities, and mass-loading factors on the evolution of the MZR.  

\section{Discussion and Conclusions}\label{sec:Conclusions}

We characterize the high-redshift gas-phase MZR in FIRE-2 simulations.  We find that the MZR from $z=5-12$ in these simulations is an approximately constant power-law relation given by $\log(Z/Z_\odot) = 0.37\log(M_*/\rm{M}_\odot) - 4.3$.  Weak evolution of the high-redshift MZR has been found in numerous other simulations (e.g., FirstLight \citep{Langan_2020}, ASTRAEUS \citep{Ucci_2023}, FLARES \citep{Wilkins_2023}), with stronger evolution being found in IllustrisTNG \citep{Torrey_2019}.  Combining our work with the analysis of the MZR in FIREbox (also run with the FIRE-2 code) from \citet{Bassini_2023}, we find that the normalization of the MZR in FIRE-2 decreases by $\sim 0.4$ dex from $z=0-3$ and evolves weakly for $z \gtrsim 3$.  

Our best-fit MZR is in good agreement with early measurements of the MZR at $z \gtrsim 5$ made by JWST.  In particular, our results are in agreement with the mean stellar mass and redshift binned results from high-redshift surveys presented by \citet{Curti_2023b} and \citet{Nakajima_2023}.  These same simulations have also been tested against the JWST UVLF by \citet{Sun_2023a, Sun_2023b}.  These agreements validate FIRE-2 as a useful predictive tool for future observations as JWST continues to expand the probed parameter region of metallicity scaling relations.

We also predict increasing scatter in the gas-phase MZR with decreasing stellar masses.  This effect may be attributable to galaxies becoming more bursty as their stellar mass decreases.  

Future work will explore the existence of the fundamental metallicity relation (FMR) in FIRE simulations.  In addition to stellar mass, the FMR suggests gas mass fraction or star formation rate as secondary predictors for metallicity (\citealp{Ellison_2008,Mannucci_2010}).  A comprehensive study on the effects of evolving gas fractions, mass loading factors, and inflow/outflow metallicities on ISM metallicities, similar to the work done by \citet{Bassini_2023} at $z = 0-3$, would be valuable to explain the evolution or lack of evolution in metallicity scaling relations at high redshift.  Beyond gas-phase metallicities, which probe the current enrichment conditions of the ISM, future work may also investigate stellar-phase metallicity scaling relations, which provide information on integrated chemical enrichment histories of galaxies.  Future analysis of the scaling relations for individual metal species tracked in FIRE simulations will allow us to make predictions for observations of individual chemical abundances made by JWST.  Finally, another emerging area of study which our simulations could inform is the measurement of metallicity gradients (radial dependence of metallicity) at high redshift.  Metallicity gradients serve as a probe for the larger processes that drive galaxy evolution in the high-redshift regime. These processes include large galactic inflows or merger events that drive bursts of star formation, generating strong feedback capable of flattening metallicity gradients, disrupting galaxy kinematics, and driving outflows.  Studying the dispersal of metals from galaxies into the intergalactic medium (IGM) will allow us to draw connections between IGM metallicity and galaxies during the epoch of reionization probed by JWST \citep{Bordoloi_2023}.  

\section*{Acknowledgements}

The authors thank Allison Strom and Luigi Bassini for useful discussions.  GS was supported by a CIERA Postdoctoral Fellowship.  CAFG was supported by NSF through grants AST-2108230, AST-2307327, and CAREER award AST-1652522; by NASA through grants 17-ATP17-0067 and 21-ATP21-0036; by STScI through grants HST-GO-16730.016-A and JWST-AR-03252.001-A; and by CXO through grant TM2-23005X. The Flatiron Institute is supported by the Simons Foundation.  RF acknowledges financial support from the Swiss National Science Foundation (grant nos PP00P2-194814 and 200021-188552). The simulations analyzed in this work were run on XSEDE computational resources (allocations TG-AST120025, TG-AST130039, TG-AST140023, and TG-AST140064).  Analysis was done using the Quest computing cluster at Northwestern University.

\vspace{5mm}

\bibliography{sample631}{}

\begin{thebibliography}{}
\expandafter\ifx\csname natexlab\endcsname\relax\def\natexlab#1{#1}\fi
\providecommand{\url}[1]{\href{#1}{#1}}
\providecommand{\dodoi}[1]{doi:~\href{http://doi.org/#1}{\nolinkurl{#1}}}
\providecommand{\doeprint}[1]{\href{http://ascl.net/#1}{\nolinkurl{http://ascl.net/#1}}}
\providecommand{\doarXiv}[1]{\href{https://arxiv.org/abs/#1}{\nolinkurl{https://arxiv.org/abs/#1}}}

\bibitem[{{Anders} \& {Grevesse}(1989)}]{Anders_1989}
{Anders}, E., \& {Grevesse}, N. 1989, \gca, 53, 197, \dodoi{10.1016/0016-7037(89)90286-X}

\bibitem[{{Bassini} {et~al.}(2024){Bassini}, {Feldmann}, {Gensior}, \& {Faucher-Giguère}}]{Bassini_2023}
{Bassini}, L., {Feldmann}, R., {Gensior}, J., \& {Faucher-Giguère}, C.~A. 2024, \mnras

\bibitem[{{Bordoloi} {et~al.}(2023){Bordoloi}, {Simcoe}, {Matthee}, {Kashino}, {Mackenzie}, {Lilly}, {Eilers}, {Liu}, {DePalma}, {Yue}, \& {Naidu}}]{Bordoloi_2023}
{Bordoloi}, R., {Simcoe}, R.~A., {Matthee}, J., {et~al.} 2023, arXiv e-prints, arXiv:2307.01273, \dodoi{10.48550/arXiv.2307.01273}

\bibitem[{{Bryan} \& {Norman}(1998)}]{Bryan_1998}
{Bryan}, G.~L., \& {Norman}, M.~L. 1998, \apj, 495, 80, \dodoi{10.1086/305262}

\bibitem[{{Bunker} {et~al.}(2023){Bunker}, {Saxena}, {Cameron}, {Willott}, {Curtis-Lake}, {Jakobsen}, {Carniani}, {Smit}, {Maiolino}, {Witstok}, {Curti}, {D'Eugenio}, {Jones}, {Ferruit}, {Arribas}, {Charlot}, {Chevallard}, {Giardino}, {de Graaff}, {Looser}, {Luetzgendorf}, {Maseda}, {Rawle}, {Rix}, {Rodriguez Del Pino}, {Alberts}, {Egami}, {Eisenstein}, {Endsley}, {Hainline}, {Hausen}, {Johnson}, {Rieke}, {Rieke}, {Robertson}, {Shivaei}, {Stark}, {Sun}, {Tacchella}, {Tang}, {Williams}, {Willmer}, {Baker}, {Baum}, {Bhatawdekar}, {Bowler}, {Boyett}, {Chen}, {Circosta}, {Helton}, {Ji}, {Lyu}, {Nelson}, {Parlanti}, {Perna}, {Sandles}, {Scholtz}, {Suess}, {Topping}, {Uebler}, {Wallace}, \& {Whitler}}]{Bunker_2023}
{Bunker}, A.~J., {Saxena}, A., {Cameron}, A.~J., {et~al.} 2023, arXiv e-prints, arXiv:2302.07256, \dodoi{10.48550/arXiv.2302.07256}

\bibitem[{{Colbrook} {et~al.}(2017){Colbrook}, {Ma}, {Hopkins}, \& {Squire}}]{Colbrook_2017}
{Colbrook}, M.~J., {Ma}, X., {Hopkins}, P.~F., \& {Squire}, J. 2017, \mnras, 467, 2421, \dodoi{10.1093/mnras/stx261}

\bibitem[{{Curti} {et~al.}(2020){Curti}, {Mannucci}, {Cresci}, \& {Maiolino}}]{Curti_2020}
{Curti}, M., {Mannucci}, F., {Cresci}, G., \& {Maiolino}, R. 2020, \mnras, 491, 944, \dodoi{10.1093/mnras/stz2910}

\bibitem[{{Curti} {et~al.}(2023{\natexlab{a}}){Curti}, {Maiolino}, {Carniani}, {D'Eugenio}, {Chevallard}, {Curtis-Lake}, {Looser}, {Scholtz}, {{\"U}bler}, {Witstok}, {Cameron}, {Charlot}, {Laseter}, {Sandles}, {Arribas}, {Bunker}, {Giardino}, {Maseda}, {Rawle}, {Rodr{\'\i}guez Del Pino}, {Smit}, {Willott}, {Eisenstein}, {Hausen}, {Johnson}, {Rieke}, {Robertson}, {Tacchella}, {Williams}, {Willmer}, {Baker}, {Bhatawdekar}, {Boyett}, {Egami}, {Helton}, {Ji}, {Kumari}, {Shivaei}, \& {Sun}}]{Curti_2023b}
{Curti}, M., {Maiolino}, R., {Carniani}, S., {et~al.} 2023{\natexlab{a}}, arXiv e-prints, arXiv:2304.08516, \dodoi{10.48550/arXiv.2304.08516}

\bibitem[{{Curti} {et~al.}(2023{\natexlab{b}}){Curti}, {D'Eugenio}, {Carniani}, {Maiolino}, {Sandles}, {Witstok}, {Baker}, {Bennett}, {Piotrowska}, {Tacchella}, {Charlot}, {Nakajima}, {Maheson}, {Mannucci}, {Amiri}, {Arribas}, {Belfiore}, {Bonaventura}, {Bunker}, {Chevallard}, {Cresci}, {Curtis-Lake}, {Hayden-Pawson}, {Jones}, {Kumari}, {Laseter}, {Looser}, {Marconi}, {Maseda}, {Scholtz}, {Smit}, {{\"U}bler}, \& {Wallace}}]{Curti_2023a}
{Curti}, M., {D'Eugenio}, F., {Carniani}, S., {et~al.} 2023{\natexlab{b}}, \mnras, 518, 425, \dodoi{10.1093/mnras/stac2737}

\bibitem[{{Ellison} {et~al.}(2008){Ellison}, {Patton}, {Simard}, \& {McConnachie}}]{Ellison_2008}
{Ellison}, S.~L., {Patton}, D.~R., {Simard}, L., \& {McConnachie}, A.~W. 2008, \apjl, 672, L107, \dodoi{10.1086/527296}

\bibitem[{{Erb} {et~al.}(2006){Erb}, {Shapley}, {Pettini}, {Steidel}, {Reddy}, \& {Adelberger}}]{Erb_2006}
{Erb}, D.~K., {Shapley}, A.~E., {Pettini}, M., {et~al.} 2006, \apj, 644, 813, \dodoi{10.1086/503623}

\bibitem[{{Escala} {et~al.}(2018){Escala}, {Wetzel}, {Kirby}, {Hopkins}, {Ma}, {Wheeler}, {Kere{\v{s}}}, {Faucher-Gigu{\`e}re}, \& {Quataert}}]{Escala_2018}
{Escala}, I., {Wetzel}, A., {Kirby}, E.~N., {et~al.} 2018, \mnras, 474, 2194, \dodoi{10.1093/mnras/stx2858}

\bibitem[{{Feldmann} {et~al.}(2023){Feldmann}, {Quataert}, {Faucher-Gigu{\`e}re}, {Hopkins}, {{\c{C}}atmabacak}, {Kere{\v{s}}}, {Bassini}, {Bernardini}, {Bullock}, {Cenci}, {Gensior}, {Liang}, {Moreno}, \& {Wetzel}}]{Feldmann_2023}
{Feldmann}, R., {Quataert}, E., {Faucher-Gigu{\`e}re}, C.-A., {et~al.} 2023, \mnras, 522, 3831, \dodoi{10.1093/mnras/stad1205}

\bibitem[{{Gill} {et~al.}(2004){Gill}, {Knebe}, \& {Gibson}}]{Gill_2004}
{Gill}, S. P.~D., {Knebe}, A., \& {Gibson}, B.~K. 2004, \mnras, 351, 399, \dodoi{10.1111/j.1365-2966.2004.07786.x}

\bibitem[{{Guo} {et~al.}(2016){Guo}, {Koo}, {Lu}, {Forbes}, {Rafelski}, {Trump}, {Amor{\'\i}n}, {Barro}, {Dav{\'e}}, {Faber}, {Hathi}, {Yesuf}, {Cooper}, {Dekel}, {Guhathakurta}, {Kirby}, {Koekemoer}, {P{\'e}rez-Gonz{\'a}lez}, {Lin}, {Newman}, {Primack}, {Rosario}, {Willmer}, \& {Yan}}]{Guo_2016}
{Guo}, Y., {Koo}, D.~C., {Lu}, Y., {et~al.} 2016, \apj, 822, 103, \dodoi{10.3847/0004-637X/822/2/103}

\bibitem[{{Henry} {et~al.}(2013{\natexlab{a}}){Henry}, {Martin}, {Finlator}, \& {Dressler}}]{Henry_2013a}
{Henry}, A., {Martin}, C.~L., {Finlator}, K., \& {Dressler}, A. 2013{\natexlab{a}}, \apj, 769, 148, \dodoi{10.1088/0004-637X/769/2/148}

\bibitem[{{Henry} {et~al.}(2013{\natexlab{b}}){Henry}, {Scarlata}, {Dom{\'\i}nguez}, {Malkan}, {Martin}, {Siana}, {Atek}, {Bedregal}, {Colbert}, {Rafelski}, {Ross}, {Teplitz}, {Bunker}, {Dressler}, {Hathi}, {Masters}, {McCarthy}, \& {Straughn}}]{Henry_2013b}
{Henry}, A., {Scarlata}, C., {Dom{\'\i}nguez}, A., {et~al.} 2013{\natexlab{b}}, \apjl, 776, L27, \dodoi{10.1088/2041-8205/776/2/L27}

\bibitem[{{Hopkins}(2015)}]{Hopkins_2015}
{Hopkins}, P.~F. 2015, \mnras, 450, 53, \dodoi{10.1093/mnras/stv195}

\bibitem[{{Hopkins} {et~al.}(2014){Hopkins}, {Kere{\v{s}}}, {O{\~n}orbe}, {Faucher-Gigu{\`e}re}, {Quataert}, {Murray}, \& {Bullock}}]{Hopkins_2014}
{Hopkins}, P.~F., {Kere{\v{s}}}, D., {O{\~n}orbe}, J., {et~al.} 2014, \mnras, 445, 581, \dodoi{10.1093/mnras/stu1738}

\bibitem[{{Hopkins} {et~al.}(2018){Hopkins}, {Wetzel}, {Kere{\v{s}}}, {Faucher-Gigu{\`e}re}, {Quataert}, {Boylan-Kolchin}, {Murray}, {Hayward}, {Garrison-Kimmel}, {Hummels}, {Feldmann}, {Torrey}, {Ma}, {Angl{\'e}s-Alc{\'a}zar}, {Su}, {Orr}, {Schmitz}, {Escala}, {Sanderson}, {Grudi{\'c}}, {Hafen}, {Kim}, {Fitts}, {Bullock}, {Wheeler}, {Chan}, {Elbert}, \& {Narayanan}}]{Hopkins_2018}
{Hopkins}, P.~F., {Wetzel}, A., {Kere{\v{s}}}, D., {et~al.} 2018, \mnras, 480, 800, \dodoi{10.1093/mnras/sty1690}

\bibitem[{{Hopkins} {et~al.}(2023){Hopkins}, {Wetzel}, {Wheeler}, {Sanderson}, {Grudi{\'c}}, {Sameie}, {Boylan-Kolchin}, {Orr}, {Ma}, {Faucher-Gigu{\`e}re}, {Kere{\v{s}}}, {Quataert}, {Su}, {Moreno}, {Feldmann}, {Bullock}, {Loebman}, {Angl{\'e}s-Alc{\'a}zar}, {Stern}, {Necib}, {Choban}, \& {Hayward}}]{Hopkins_2023}
{Hopkins}, P.~F., {Wetzel}, A., {Wheeler}, C., {et~al.} 2023, \mnras, 519, 3154, \dodoi{10.1093/mnras/stac3489}

\bibitem[{{Hsiao} {et~al.}(2024){Hsiao}, {{\'A}lvarez-M{\'a}rquez}, {Coe}, {Crespo G{\'o}mez}, {Abdurro'uf}, {Dayal}, {Larson}, {Bik}, {Blanco-Prieto}, {Colina}, {P{\'e}rez-Gonz{\'a}lez}, {Costantin}, {Prieto-Jim{\'e}nez}, {Adamo}, {Bradley}, {Conselice}, {Fujimoto}, {Furtak}, {Hutchison}, {James}, {Jim{\'e}nez-Teja}, {Jung}, {Kokorev}, {Mingozzi}, {Norman}, {Ricotti}, {Rigby}, {Sharon}, {Vanzella}, {Welch}, {Xu}, {Zackrisson}, \& {Zitrin}}]{Hsiao_2024}
{Hsiao}, T. Y.-Y., {{\'A}lvarez-M{\'a}rquez}, J., {Coe}, D., {et~al.} 2024, arXiv e-prints, arXiv:2404.16200.
\newblock \doarXiv{2404.16200}

\bibitem[{{Knollmann} \& {Knebe}(2009)}]{Knollmann_2009}
{Knollmann}, S.~R., \& {Knebe}, A. 2009, \apjs, 182, 608, \dodoi{10.1088/0067-0049/182/2/608}

\bibitem[{{Kroupa}(2001)}]{Kroupa_2001}
{Kroupa}, P. 2001, \mnras, 322, 231, \dodoi{10.1046/j.1365-8711.2001.04022.x}

\bibitem[{{Langan} {et~al.}(2020){Langan}, {Ceverino}, \& {Finlator}}]{Langan_2020}
{Langan}, I., {Ceverino}, D., \& {Finlator}, K. 2020, \mnras, 494, 1988, \dodoi{10.1093/mnras/staa880}

\bibitem[{{Lee} {et~al.}(2006){Lee}, {Skillman}, {Cannon}, {Jackson}, {Gehrz}, {Polomski}, \& {Woodward}}]{Lee_2006}
{Lee}, H., {Skillman}, E.~D., {Cannon}, J.~M., {et~al.} 2006, \apj, 647, 970, \dodoi{10.1086/505573}

\bibitem[{{Leitherer} {et~al.}(1999){Leitherer}, {Schaerer}, {Goldader}, {Delgado}, {Robert}, {Kune}, {de Mello}, {Devost}, \& {Heckman}}]{Leitherer_1999}
{Leitherer}, C., {Schaerer}, D., {Goldader}, J.~D., {et~al.} 1999, \apjs, 123, 3, \dodoi{10.1086/313233}

\bibitem[{{Lequeux} {et~al.}(1979){Lequeux}, {Peimbert}, {Rayo}, {Serrano}, \& {Torres-Peimbert}}]{1979A&A....80..155L}
{Lequeux}, J., {Peimbert}, M., {Rayo}, J.~F., {Serrano}, A., \& {Torres-Peimbert}, S. 1979, \aap, 80, 155

\bibitem[{{Li} {et~al.}(2023){Li}, {Cai}, {Bian}, {Lin}, {Li}, {Wu}, {Sun}, {Zhang}, {Golden-Marx}, {Sun}, {Zou}, {Fan}, {Egami}, {Charlot}, {Bruzual}, \& {Chevallard}}]{Li_2023}
{Li}, M., {Cai}, Z., {Bian}, F., {et~al.} 2023, \apjl, 955, L18, \dodoi{10.3847/2041-8213/acf470}

\bibitem[{{Lilly} {et~al.}(2013){Lilly}, {Carollo}, {Pipino}, {Renzini}, \& {Peng}}]{Lilly_2013}
{Lilly}, S.~J., {Carollo}, C.~M., {Pipino}, A., {Renzini}, A., \& {Peng}, Y. 2013, \apj, 772, 119, \dodoi{10.1088/0004-637X/772/2/119}

\bibitem[{{Ma} {et~al.}(2016){Ma}, {Hopkins}, {Faucher-Gigu{\`e}re}, {Zolman}, {Muratov}, {Kere{\v{s}}}, \& {Quataert}}]{Ma_2016}
{Ma}, X., {Hopkins}, P.~F., {Faucher-Gigu{\`e}re}, C.-A., {et~al.} 2016, \mnras, 456, 2140, \dodoi{10.1093/mnras/stv2659}

\bibitem[{{Ma} {et~al.}(2018{\natexlab{a}}){Ma}, {Hopkins}, {Boylan-Kolchin}, {Faucher-Gigu{\`e}re}, {Quataert}, {Feldmann}, {Garrison-Kimmel}, {Hayward}, {Kere{\v{s}}}, \& {Wetzel}}]{Ma_2018a}
{Ma}, X., {Hopkins}, P.~F., {Boylan-Kolchin}, M., {et~al.} 2018{\natexlab{a}}, \mnras, 477, 219, \dodoi{10.1093/mnras/sty684}

\bibitem[{{Ma} {et~al.}(2018{\natexlab{b}}){Ma}, {Hopkins}, {Garrison-Kimmel}, {Faucher-Gigu{\`e}re}, {Quataert}, {Boylan-Kolchin}, {Hayward}, {Feldmann}, \& {Kere{\v{s}}}}]{Ma_2018b}
{Ma}, X., {Hopkins}, P.~F., {Garrison-Kimmel}, S., {et~al.} 2018{\natexlab{b}}, \mnras, 478, 1694, \dodoi{10.1093/mnras/sty1024}

\bibitem[{{Ma} {et~al.}(2019){Ma}, {Hayward}, {Casey}, {Hopkins}, {Quataert}, {Liang}, {Faucher-Gigu{\`e}re}, {Feldmann}, \& {Kere{\v{s}}}}]{Ma_2019}
{Ma}, X., {Hayward}, C.~C., {Casey}, C.~M., {et~al.} 2019, \mnras, 487, 1844, \dodoi{10.1093/mnras/stz1324}

\bibitem[{{Maier} {et~al.}(2014){Maier}, {Lilly}, {Ziegler}, {Contini}, {P{\'e}rez Montero}, {Peng}, \& {Balestra}}]{Maier_2014}
{Maier}, C., {Lilly}, S.~J., {Ziegler}, B.~L., {et~al.} 2014, \apj, 792, 3, \dodoi{10.1088/0004-637X/792/1/3}

\bibitem[{{Maiolino} {et~al.}(2008){Maiolino}, {Nagao}, {Grazian}, {Cocchia}, {Marconi}, {Mannucci}, {Cimatti}, {Pipino}, {Fontana}, {Granato}, {Matteucci}, {Pentericci}, {Risaliti}, {Salvati}, \& {Silva}}]{Maiolino_2008}
{Maiolino}, R., {Nagao}, T., {Grazian}, A., {et~al.} 2008, in Astronomical Society of the Pacific Conference Series, Vol. 396, Formation and Evolution of Galaxy Disks, ed. J.~G. {Funes} \& E.~M. {Corsini}, 409

\bibitem[{{Mannucci} {et~al.}(2010){Mannucci}, {Cresci}, {Maiolino}, {Marconi}, \& {Gnerucci}}]{Mannucci_2010}
{Mannucci}, F., {Cresci}, G., {Maiolino}, R., {Marconi}, A., \& {Gnerucci}, A. 2010, \mnras, 408, 2115, \dodoi{10.1111/j.1365-2966.2010.17291.x}

\bibitem[{{Mannucci} {et~al.}(2009){Mannucci}, {Cresci}, {Maiolino}, {Marconi}, {Pastorini}, {Pozzetti}, {Gnerucci}, {Risaliti}, {Schneider}, {Lehnert}, \& {Salvati}}]{Mannucci_2009}
{Mannucci}, F., {Cresci}, G., {Maiolino}, R., {et~al.} 2009, \mnras, 398, 1915, \dodoi{10.1111/j.1365-2966.2009.15185.x}

\bibitem[{{Marszewski} {et~al.}(in prep.){Marszewski}, {Sun}, \& {Faucher-Gigu{\`e}re}}]{Marszewski_inprep}
{Marszewski}, A., {Sun}, G., \& {Faucher-Gigu{\`e}re}, C.-A. in prep.

\bibitem[{{Morishita} {et~al.}(2024){Morishita}, {Stiavelli}, {Grillo}, {Rosati}, {Schuldt}, {Trenti}, {Bergamini}, {Boyett}, {Chary}, {Leethochawalit}, {Roberts-Borsani}, {Treu}, \& {Vanzella}}]{Morishita_2024}
{Morishita}, T., {Stiavelli}, M., {Grillo}, C., {et~al.} 2024, arXiv e-prints, arXiv:2402.14084, \dodoi{10.48550/arXiv.2402.14084}

\bibitem[{{Nakajima} {et~al.}(2023){Nakajima}, {Ouchi}, {Isobe}, {Harikane}, {Zhang}, {Ono}, {Umeda}, \& {Oguri}}]{Nakajima_2023}
{Nakajima}, K., {Ouchi}, M., {Isobe}, Y., {et~al.} 2023, arXiv e-prints, arXiv:2301.12825, \dodoi{10.48550/arXiv.2301.12825}

\bibitem[{{Pallottini} {et~al.}(2022){Pallottini}, {Ferrara}, {Gallerani}, {Behrens}, {Kohandel}, {Carniani}, {Vallini}, {Salvadori}, {Gelli}, {Sommovigo}, {D'Odorico}, {Di Mascia}, \& {Pizzati}}]{Pallottini_2022}
{Pallottini}, A., {Ferrara}, A., {Gallerani}, S., {et~al.} 2022, \mnras, 513, 5621, \dodoi{10.1093/mnras/stac1281}

\bibitem[{{Planck Collaboration} {et~al.}(2020){Planck Collaboration}, {Aghanim}, {Akrami}, {Ashdown}, {Aumont}, {Baccigalupi}, {Ballardini}, {Banday}, {Barreiro}, {Bartolo}, {Basak}, {Battye}, {Benabed}, {Bernard}, {Bersanelli}, {Bielewicz}, {Bock}, {Bond}, {Borrill}, {Bouchet}, {Boulanger}, {Bucher}, {Burigana}, {Butler}, {Calabrese}, {Cardoso}, {Carron}, {Challinor}, {Chiang}, {Chluba}, {Colombo}, {Combet}, {Contreras}, {Crill}, {Cuttaia}, {de Bernardis}, {de Zotti}, {Delabrouille}, {Delouis}, {Di Valentino}, {Diego}, {Dor{\'e}}, {Douspis}, {Ducout}, {Dupac}, {Dusini}, {Efstathiou}, {Elsner}, {En{\ss}lin}, {Eriksen}, {Fantaye}, {Farhang}, {Fergusson}, {Fernandez-Cobos}, {Finelli}, {Forastieri}, {Frailis}, {Fraisse}, {Franceschi}, {Frolov}, {Galeotta}, {Galli}, {Ganga}, {G{\'e}nova-Santos}, {Gerbino}, {Ghosh}, {Gonz{\'a}lez-Nuevo}, {G{\'o}rski}, {Gratton}, {Gruppuso}, {Gudmundsson}, {Hamann}, {Handley}, {Hansen}, {Herranz}, {Hildebrandt}, {Hivon}, {Huang}, {Jaffe}, {Jones}, {Karakci}, {Keih{\"a}nen},
  {Keskitalo}, {Kiiveri}, {Kim}, {Kisner}, {Knox}, {Krachmalnicoff}, {Kunz}, {Kurki-Suonio}, {Lagache}, {Lamarre}, {Lasenby}, {Lattanzi}, {Lawrence}, {Le Jeune}, {Lemos}, {Lesgourgues}, {Levrier}, {Lewis}, {Liguori}, {Lilje}, {Lilley}, {Lindholm}, {L{\'o}pez-Caniego}, {Lubin}, {Ma}, {Mac{\'\i}as-P{\'e}rez}, {Maggio}, {Maino}, {Mandolesi}, {Mangilli}, {Marcos-Caballero}, {Maris}, {Martin}, {Martinelli}, {Mart{\'\i}nez-Gonz{\'a}lez}, {Matarrese}, {Mauri}, {McEwen}, {Meinhold}, {Melchiorri}, {Mennella}, {Migliaccio}, {Millea}, {Mitra}, {Miville-Desch{\^e}nes}, {Molinari}, {Montier}, {Morgante}, {Moss}, {Natoli}, {N{\o}rgaard-Nielsen}, {Pagano}, {Paoletti}, {Partridge}, {Patanchon}, {Peiris}, {Perrotta}, {Pettorino}, {Piacentini}, {Polastri}, {Polenta}, {Puget}, {Rachen}, {Reinecke}, {Remazeilles}, {Renzi}, {Rocha}, {Rosset}, {Roudier}, {Rubi{\~n}o-Mart{\'\i}n}, {Ruiz-Granados}, {Salvati}, {Sandri}, {Savelainen}, {Scott}, {Shellard}, {Sirignano}, {Sirri}, {Spencer}, {Sunyaev}, {Suur-Uski}, {Tauber}, {Tavagnacco},
  {Tenti}, {Toffolatti}, {Tomasi}, {Trombetti}, {Valenziano}, {Valiviita}, {Van Tent}, {Vibert}, {Vielva}, {Villa}, {Vittorio}, {Wandelt}, {Wehus}, {White}, {White}, {Zacchei}, \& {Zonca}}]{Planck_2020}
{Planck Collaboration}, {Aghanim}, N., {Akrami}, Y., {et~al.} 2020, \aap, 641, A6, \dodoi{10.1051/0004-6361/201833910}

\bibitem[{{Sanders} {et~al.}(2015){Sanders}, {Shapley}, {Kriek}, {Reddy}, {Freeman}, {Coil}, {Siana}, {Mobasher}, {Shivaei}, {Price}, \& {de Groot}}]{Sanders_2015}
{Sanders}, R.~L., {Shapley}, A.~E., {Kriek}, M., {et~al.} 2015, \apj, 799, 138, \dodoi{10.1088/0004-637X/799/2/138}

\bibitem[{{Steidel} {et~al.}(2014){Steidel}, {Rudie}, {Strom}, {Pettini}, {Reddy}, {Shapley}, {Trainor}, {Erb}, {Turner}, {Konidaris}, {Kulas}, {Mace}, {Matthews}, \& {McLean}}]{Steidel_2014}
{Steidel}, C.~C., {Rudie}, G.~C., {Strom}, A.~L., {et~al.} 2014, \apj, 795, 165, \dodoi{10.1088/0004-637X/795/2/165}

\bibitem[{{Sun} {et~al.}(2023{\natexlab{a}}){Sun}, {Faucher-Gigu{\`e}re}, {Hayward}, \& {Shen}}]{Sun_2023a}
{Sun}, G., {Faucher-Gigu{\`e}re}, C.-A., {Hayward}, C.~C., \& {Shen}, X. 2023{\natexlab{a}}, \mnras, 526, 2665, \dodoi{10.1093/mnras/stad2902}

\bibitem[{{Sun} {et~al.}(2023{\natexlab{b}}){Sun}, {Faucher-Gigu{\`e}re}, {Hayward}, {Shen}, {Wetzel}, \& {Cochrane}}]{Sun_2023b}
{Sun}, G., {Faucher-Gigu{\`e}re}, C.-A., {Hayward}, C.~C., {et~al.} 2023{\natexlab{b}}, \apjl, 955, L35, \dodoi{10.3847/2041-8213/acf85a}

\bibitem[{{Torrey} {et~al.}(2019){Torrey}, {Vogelsberger}, {Marinacci}, {Pakmor}, {Springel}, {Nelson}, {Naiman}, {Pillepich}, {Genel}, {Weinberger}, \& {Hernquist}}]{Torrey_2019}
{Torrey}, P., {Vogelsberger}, M., {Marinacci}, F., {et~al.} 2019, \mnras, 484, 5587, \dodoi{10.1093/mnras/stz243}

\bibitem[{{Tremonti} {et~al.}(2004){Tremonti}, {Heckman}, {Kauffmann}, {Brinchmann}, {Charlot}, {White}, {Seibert}, {Peng}, {Schlegel}, {Uomoto}, {Fukugita}, \& {Brinkmann}}]{Tremonti_2004}
{Tremonti}, C.~A., {Heckman}, T.~M., {Kauffmann}, G., {et~al.} 2004, \apj, 613, 898, \dodoi{10.1086/423264}

\bibitem[{{Ucci} {et~al.}(2023){Ucci}, {Dayal}, {Hutter}, {Kobayashi}, {Gottl{\"o}ber}, {Yepes}, {Hunt}, {Legrand}, \& {Tortora}}]{Ucci_2023}
{Ucci}, G., {Dayal}, P., {Hutter}, A., {et~al.} 2023, \mnras, 518, 3557, \dodoi{10.1093/mnras/stac2654}

\bibitem[{{Wilkins} {et~al.}(2023){Wilkins}, {Vijayan}, {Lovell}, {Roper}, {Irodotou}, {Caruana}, {Seeyave}, {Kuusisto}, {Thomas}, \& {Parris}}]{Wilkins_2023}
{Wilkins}, S.~M., {Vijayan}, A.~P., {Lovell}, C.~C., {et~al.} 2023, \mnras, 519, 3118, \dodoi{10.1093/mnras/stac3280}

\bibitem[{{Yabe} {et~al.}(2014){Yabe}, {Ohta}, {Iwamuro}, {Akiyama}, {Tamura}, {Yuma}, {Kimura}, {Takato}, {Moritani}, {Sumiyoshi}, {Maihara}, {Silverman}, {Dalton}, {Lewis}, {Bonfield}, {Lee}, {Curtis-Lake}, {Macaulay}, \& {Clarke}}]{Yabe_2014}
{Yabe}, K., {Ohta}, K., {Iwamuro}, F., {et~al.} 2014, \mnras, 437, 3647, \dodoi{10.1093/mnras/stt2185}

\bibitem[{{Yang} {et~al.}(2023){Yang}, {Lidz}, {Smith}, {Benson}, \& {Li}}]{Yang_2023}
{Yang}, S., {Lidz}, A., {Smith}, A., {Benson}, A., \& {Li}, H. 2023, \mnras, 525, 5989, \dodoi{10.1093/mnras/stad2571}

\bibitem[{{Zahid} {et~al.}(2012){Zahid}, {Bresolin}, {Kewley}, {Coil}, \& {Dav{\'e}}}]{Zahid_2012}
{Zahid}, H.~J., {Bresolin}, F., {Kewley}, L.~J., {Coil}, A.~L., \& {Dav{\'e}}, R. 2012, \apj, 750, 120, \dodoi{10.1088/0004-637X/750/2/120}

\bibitem[{{Zahid} {et~al.}(2013){Zahid}, {Geller}, {Kewley}, {Hwang}, {Fabricant}, \& {Kurtz}}]{Zahid_2013}
{Zahid}, H.~J., {Geller}, M.~J., {Kewley}, L.~J., {et~al.} 2013, \apjl, 771, L19, \dodoi{10.1088/2041-8205/771/2/L19}

\bibitem[{{Zahid} {et~al.}(2011){Zahid}, {Kewley}, \& {Bresolin}}]{Zahid_2011}
{Zahid}, H.~J., {Kewley}, L.~J., \& {Bresolin}, F. 2011, \apj, 730, 137, \dodoi{10.1088/0004-637X/730/2/137}

\end{thebibliography}
\bibliographystyle{aasjournal}

\begin{appendices}

\section{Comparison with Previous FIRE Work} \label {appendix:FIRE}

\begin{figure}[h!]
    \centering
    \includegraphics[width=0.5\linewidth]{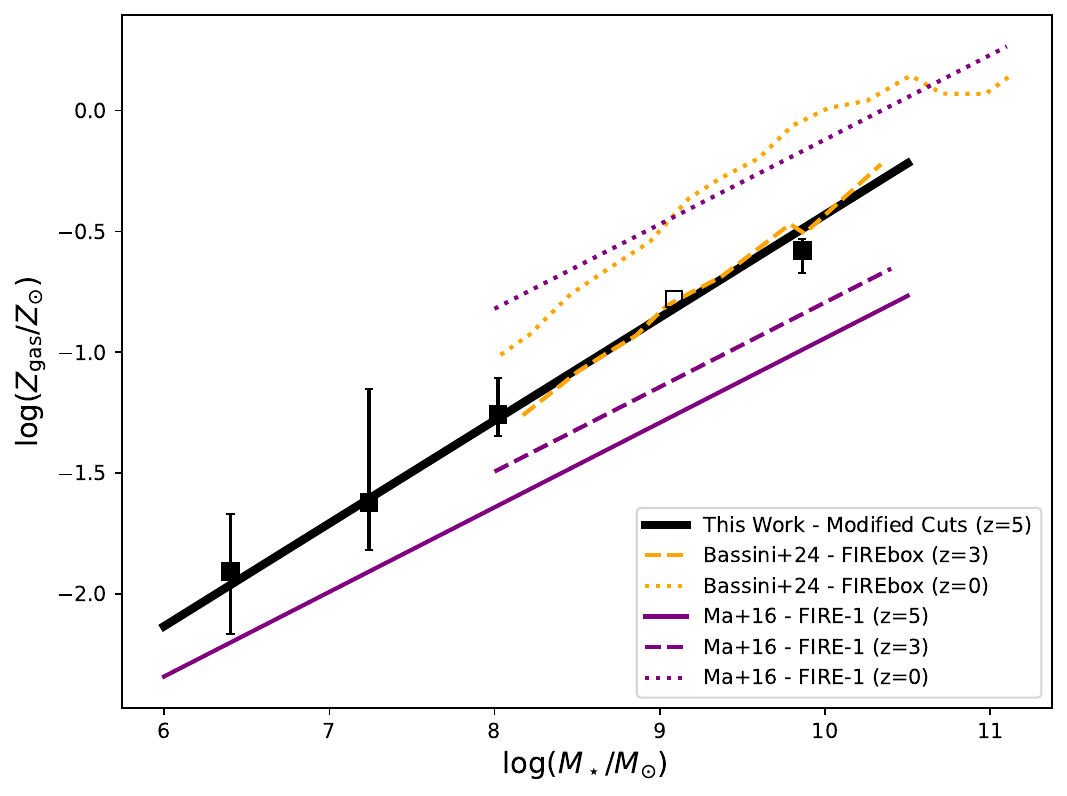}
    \caption{Comparison between the MZR at $z=5$ from the high-redshift suite of FIRE-2 simulations analyzed in this paper and other FIRE work on the MZR.  Here, temperature and radial cuts are made on the high-redshift FIRE-2 galaxies to match those made by \citet{Ma_2016} in analyzing FIRE-1 galaxies.  Our $z=5$ fit is in good agreement with the $z=3$ result from FIREbox (run with the FIRE-2 code) presented by \citet{Bassini_2023} (shown in orange), indicating weak evolution of the MZR in FIRE-2 galaxies for $z\gtrsim3$.  The fit from FIRE-1 simulations (shown in purple) remains consistently $0.3$--$0.4$ dex below this work at $z=5$ and $0.15$--$0.4$ dex below the FIREbox result at $z=3$.  The agreement between FIREbox and FIRE-1 at $z=0$ demonstrates that this offset between FIRE-1 and FIRE-2 runs 
    is only present at higher redshift.}
    \label{fig:FIRE_Comparison}
\end{figure}

In Figure \ref{fig:FIRE_Comparison}, we compare our MZR results with previous results from FIRE simulations. 
First, we compare to the MZR in FIRE-1 zoom-in simulations from \citet{Ma_2016}.
As in Figure \ref{fig:Models}, we find a significant ($\sim0.3$--$0.4$ dex) offset between the MZR in this work and that from FIRE-1 presented by \citet{Ma_2016} at $z=5$. For the comparison shown, we repeated our analysis of the MZR matching the temperature and radial cuts on the gas particles included in our analysis with those made by \citet{Ma_2016} for consistency. These cuts include all gas with temperature below $10^4$ K within $0.1R_{\rm vir}$ of the galaxy's center (in the main body of the paper, we included all gas within $0.2R_{\rm vir}$). The change in cuts on the gas does not appreciably change our results and the offset between the FIRE-1 result and this work remains.

We additionally compare this work with a recent study of the MZR by \citet{Bassini_2023} based on the FIREbox simulation, run with the FIRE-2 code like our zoom-in simulations, but analyzed over $z=0-3$.  While we apply slightly different cuts (they consider all gas within $0.1R_{\rm vir}$), we find close agreement between the MZR at our lowest redshift analyzed ($z=5$) and their highest redshift analyzed ($z=3$).  This agreement suggests the absence of significant evolution in the MZR in FIRE-2 simulations for $z \gtrsim 3$. 
The FIREbox data also allow us to compare FIRE-2 vs. FIRE-1 runs at different redshifts. 
We find that the FIRE-1 fit from \cite{Ma_2016} is significantly offset from the FIREbox result at $z=3$ ($0.15$--$0.4$ dex below). 
However, the offset between FIRE-1 and FIRE-2 largely vanishes by $z=0$. 
This is reassuring because the comparison between FIRE-1 and FIRE-2 in \cite{Hopkins_2018}, which found no major differences in the stellar mass-halo mass relation between the two sets of simulations, focused on $z=0$. 
This suggests that some aspects of the cosmic baryon cycle which determines the enrichment of galaxies differ significantly between FIRE-1 and FIRE-2 at high redshift, even though the stellar masses and galaxy metallicities converge to broadly consistent values by $z=0$. 
It is beyond the scope of this work to fully investigate the cause of this offset as FIRE-2 implemented a large number of improvements and changes that could impact the metal enrichment of the ISM as well as the driving of inflows and outflows. Changes made between the FIRE-1 and FIRE-2 codes include the introduction of a more accurate hydrodynamic solver, a supernova feedback scheme that more accurately conserves momentum, and updated metal yields (see \citealt{Hopkins_2018} for an exhaustive list and full descriptions of improvements).

\section{Evolving Gas Fractions} \label {appendix:Gas_Fraction}

\begin{figure}[h!]
    \centering
    \includegraphics[width=0.95\linewidth]{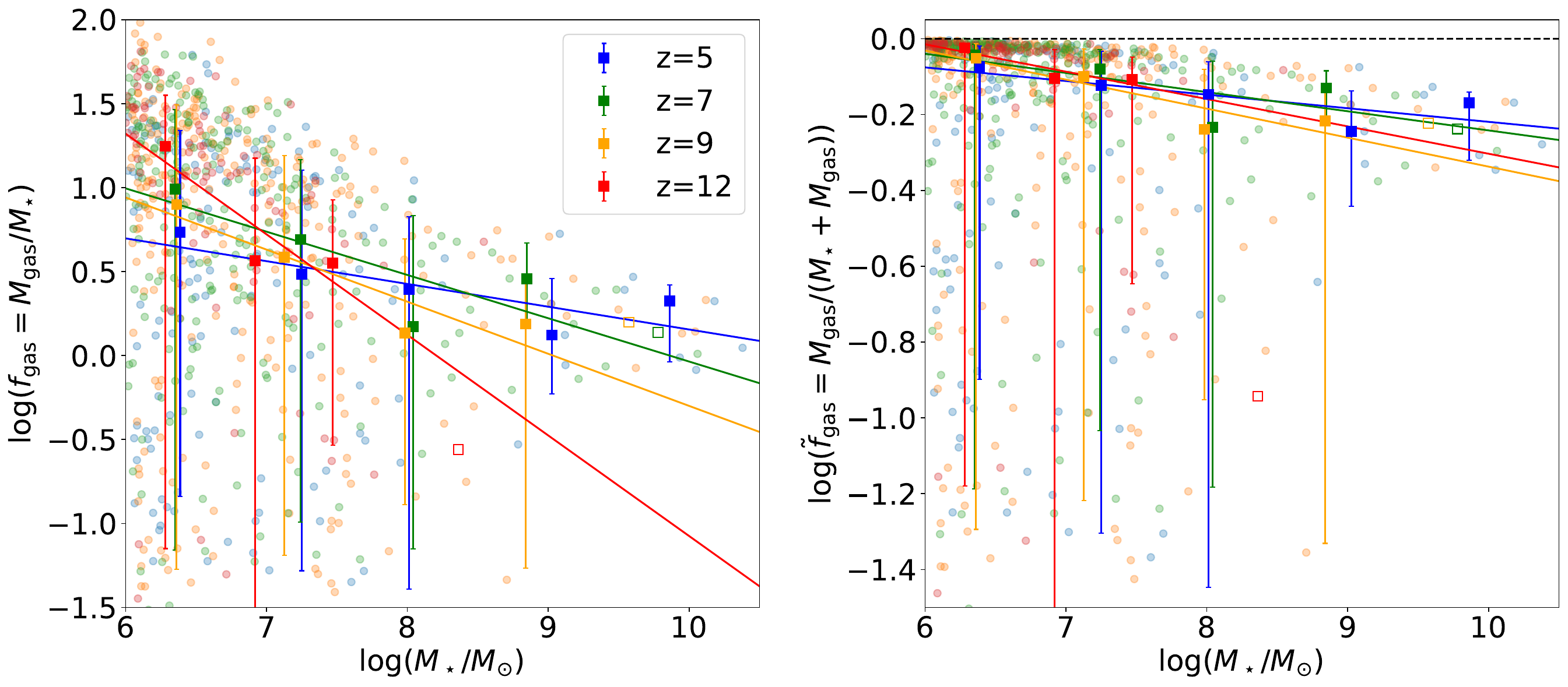}
    \caption{The gas fractions, $f_{\rm gas} = M_{\rm gas}/M_{\star}$ (left) and $\tilde{f}_{\rm gas} = M_{\rm gas}/(M_{\star}+M_{\rm gas})$ (right), in different stellar mass bins for $z=5$ (blue), $z=7$ (green), $z=9$ (yellow), and $z=12$ (red).  Individual galaxies are shown by dots.  The median gas fractions for stellar mass bins containing at least five galaxies are shown by solid squares with error bars representing the 16th and 84th percentiles.  Empty squares represent the median gas fractions of stellar mass bins that contain fewer than 5 galaxies.  Solid lines present best fits to the median gas fractions.  At some stellar masses, median gas fractions show large fluctuations between different redshifts.  The systematic redshift evolution is especially clear for the definition on the left.}
    \label{fig:Gas_Fraction}
\end{figure}

Previous works have used galaxies' gas fractions to explain the evolution of the MZR using either gas-regulator models or “closed/leaky box" models.  In the gas-regulator model we define the gas fraction to be $f_{\rm gas} = M_{\rm gas}/M_{\star}$, where $M_{\rm gas}$ and $M_{\star}$ are the total gas mass and stellar mass within $0.2R_{\rm vir}$ of the center of a galaxy, respectively.  Previous works have used the redshift evolution of galaxies' gas fractions to explain the evolution of the MZR for $z \lesssim 3.5$.  However, \citet{Bassini_2023} show that evolving gas fractions are not responsible for the evolving MZR in this redshift range in the FIREbox simulation.  Rather, the redshift dependence on the metallicities of gas inflows and outflows as well as the evolution of the mass loading factor drive the evolution of the MZR at lower redshifts.  Other works have cited weak evolution of gas fractions at high redshift as being responsible for weak evolution of the high-redshift ($z \gtrsim 3.5$) MZR (e.g., \citealp{Torrey_2019}).  The left panel of Figure \ref{fig:Gas_Fraction} presents stellar mass-binned gas fractions from our simulations that vary substantially with redshift and in a mass-dependent way, indicating that gas fractions alone cannot explain the weakly evolving high-redshift MZR.   

In the “closed box" or “leaky box" models, we define a second version of the gas fraction as $\tilde{f}_{\rm gas} = M_{\rm gas}/(M_{\star}+M_{\rm gas})$.  The right panel of Figure \ref{fig:Gas_Fraction} shows values of $\tilde{f}_{\rm gas}$ in our simulated galaxies.  The saturation of $\tilde{f}_{\rm gas}$ to unity and/or its weak evolution at high redshift has been used to explain the weak evolution in the MZR for $z \gtrsim 3.5$ (e.g., \citealp{Ma_2016,Langan_2020}).  However, both of these works find that, in order for their simulation data to be well-fit by a "leaky box" model, they must use an effective stellar yield ($y_{\rm eff}$) that is much smaller than the intrinsic stellar yield ($y=0.02$).  The significantly smaller value of $y_{\rm eff}$ implies that there is a large net impact of inflows and outflows on metallicities.  Thus, the weak evolution of the MZR at high redshift cannot be attributed to saturated or weakly evolving values of $\tilde{f}_{\rm gas}$ alone and must take into account the larger effects of inflows and outflows.  

\section{Alternative Calculations of Metallicities} \label {appendix:Gas_Cuts}

Throughout this work we define a galaxy's gas-phase metallicity to be the mass-weighted mean metallicity of all gas particles within $0.2R_{\rm vir}$ of the galaxy's center.  Here, we investigate the effects of using alternative cuts on the gas particles included in our analysis.  We consider a smaller radial cut on our galaxy ($r_{\rm gas} < 0.1R_{\rm vir}$).  We also present a version of our analysis with radial and temperature cuts that exactly match those used by \citet{Ma_2016} in their analysis of the MZR in FIRE-1 ($r_{\rm gas} < 0.1R_{\rm vir}$ and $T_{\rm gas} < 10^4$ K).  Applying either of these cuts does not appear to have a significant effect on our resulting MZR.  We ultimately elect to use $r_{\rm gas} < 0.2R_{\rm vir}$ as our radial cut due to the tendency of high-redshift galaxies to have more spatially extended stellar populations relative to their virial radii as compared to galaxies at lower redshift.  We also choose to not include a temperature cut on our gas as this cut would eliminate a significant number of galaxies from our sample.

\begin{figure}[h!]
    \centering
    \includegraphics[width=0.95\linewidth]{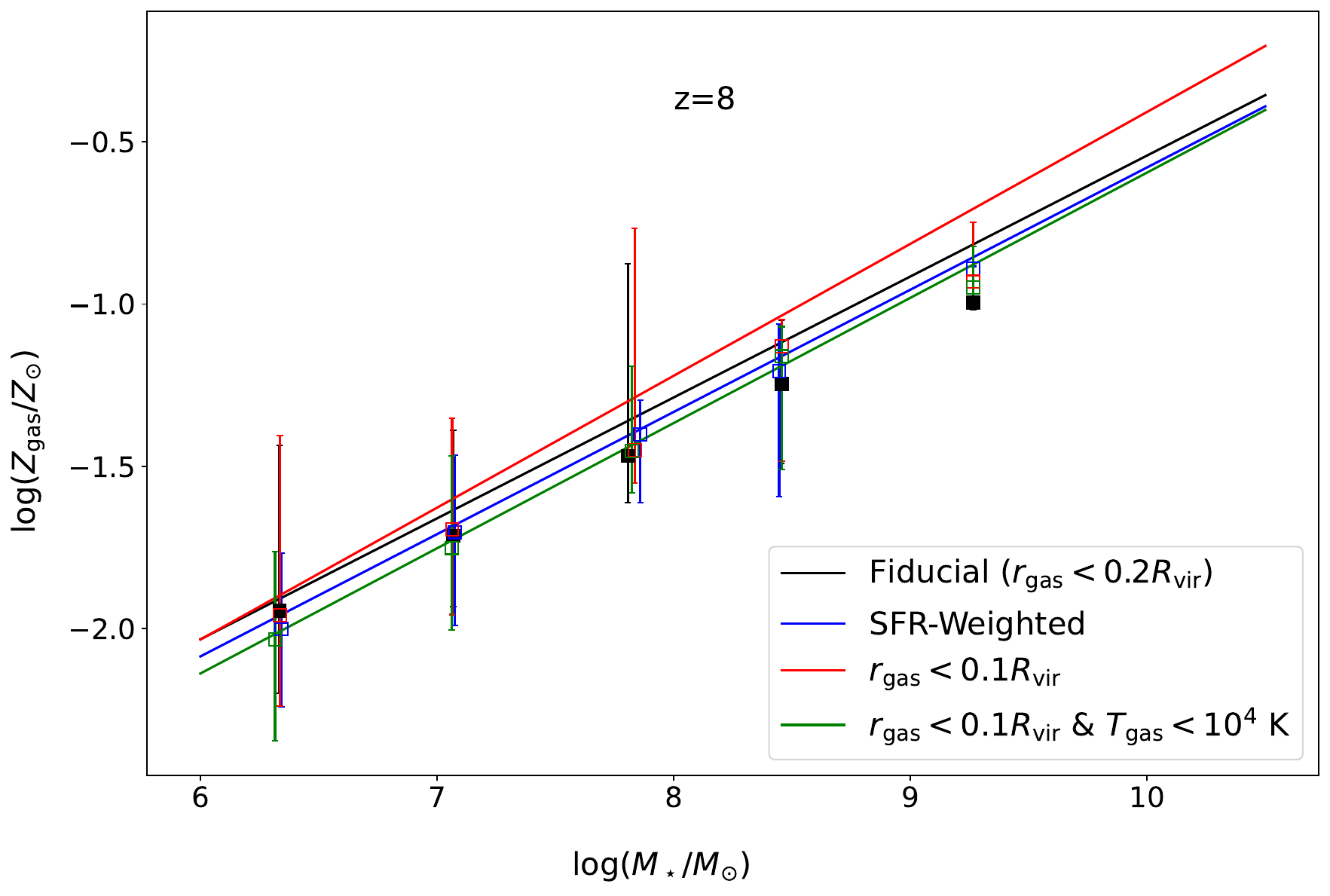}
    \caption{The gas-phase MZR in our simulations at $z=8$ calculated using different weighting schemes and cuts on the gas included.  The solid lines show our best fits to the medians, allowing for redshift evolution of the slope and normalization.  The median metallicities of stellar mass bins are shown by squares with error bars representing the 16th and 84th percentiles.  Note that, while the medians are shown at $z=8$ only, the fits are calculated across all redshifts analyzed ($z = 5-12$), and thus, may be offset from the medians shown. Our fiducial method, where metallicity is calculated as the mass-weighted mean metallicity of all gas particles within $0.2R_{\rm vir}$, is shown in black.  The method where metallicities are calculated as the SFR-weighted mean metallicity of all gas particles within $0.2R_{\rm vir}$, is shown in blue.  The mass-weighted method only considering gas particles within $0.1R_{\rm vir}$ is shown in red.  The mass-weighted method including a temperature cut ($T_{\rm gas} < 10^4$ K) and only considering gas particles within $0.1R_{\rm vir}$ is shown in green. 
 Different weighting schemes and cuts on the gas do not appear to significantly change our resulting MZR with all models differing from our fiducial model by $\lesssim 0.1$ dex across our stellar mass range.}
    \label{fig:Alternative_Cuts}
\end{figure}

We also consider weighting gas particles by their star formation rate property rather than by their mass when calculating a galaxy's metallicity.  This SFR-weighting scheme does not have a significant impact on our calculated MZR and likely introduces a bias by removing galaxies with no star-forming gas from our sample.  We therefore elect to use mass-weighted metallicities.  A comparison between the MZR calculated using our fiducial method and the MZR calculated using the alternative cuts and the SFR-weighting method is shown in Figure \ref{fig:Alternative_Cuts}.  

\end{appendices}

\end{document}